\documentclass[twocolumn,aps,pra]{revtex4-1}

\usepackage{ucs}
\usepackage[utf8x]{inputenc}
\usepackage[T1]{fontenc}
\usepackage[american]{babel}

\usepackage{lmodern}

\usepackage{isotope}
\usepackage{textcomp,gensymb}
\usepackage{amssymb,amsmath}
\usepackage{units}
\usepackage{bm}

\newcommand*{\dd}{\ensuremath{\mathrm{d}}}
\newcommand*{\ii}{\ensuremath{\mathrm{i}}}
\newcommand*{\ee}{\ensuremath{\mathrm{e}}}
\newcommand*{\vc}[1]{\ensuremath{\bm{#1}}}
\newcommand*{\unitvc}[1]{\ensuremath{\bm{\hat #1}}}
\newcommand*{\unitx}{\unitvc{x}}

\newcommand*{\abs}[1]{\ensuremath{\left\lvert#1\right\rvert}}
\newcommand*{\avg}[1]{\ensuremath{\left\langle{#1}\right\rangle}}

\newcommand*{\var}[1]{\ensuremath{\avg{\Delta #1^2}}}

\newcommand*{\pv}{\ensuremath{\mathrm{P}}}

\newcommand*{\bra}[1]{\ensuremath{\left\langle#1\right\rvert}}
\newcommand*{\ket}[1]{\ensuremath{\left\lvert#1\right\rangle}}

\newcommand*{\comm}[2]{\ensuremath{\left[#1,#2\right]}}

\newcommand{\adj}[1]{\ensuremath{#1^\dagger}}

\newcommand*{\upk}{\ensuremath{\ket{\uparrow}}}
\newcommand*{\upb}{\ensuremath{\bra{\uparrow}}}
\newcommand*{\downk}{\ensuremath{\ket{\downarrow}}}
\newcommand*{\downb}{\ensuremath{\bra{\downarrow}}}
\newcommand*{\Sx}{\ensuremath{S_x}}
\newcommand*{\Sy}{\ensuremath{S_y}}
\newcommand*{\Sz}{\ensuremath{S_z}}
\newcommand*{\Splus}{\ensuremath{S_+}}
\newcommand*{\Sminus}{\ensuremath{S_-}}

\newcommand*{\Rb}{\isotope[87]{Rb}}

\newcommand*{\exck}{\ensuremath{\ket{e}}}
\newcommand*{\excb}{\ensuremath{\bra{e}}}
\newcommand*{\ocav}{\ensuremath{\omega_c}}
\newcommand*{\oat}{\ensuremath{\omega_a}}
\newcommand*{\oprobe}{\ensuremath{{\omega_p}}}
\newcommand*{\afield}[1]{\ensuremath{a_{#1}}}
\newcommand*{\adag}[1]{\ensuremath{\adj{a}_{#1}}}
\newcommand*{\na}[1]{\ensuremath{\adag{#1}\afield{#1}}}
\newcommand*{\bfield}[1]{\ensuremath{b_{#1}}}
\newcommand*{\bdag}[1]{\ensuremath{\adj{b}_{#1}}}
\newcommand*{\nb}[1]{\ensuremath{\bdag{#1}\bfield{#1}}}
\newcommand*{\cfield}{\ensuremath{c}}
\newcommand*{\cdag}{\ensuremath{\adj{c}}}
\newcommand*{\nc}{\ensuremath{\cdag \cfield}}
\newcommand*{\pphase}{\ensuremath{\phi_0}}
\newcommand*{\Smin}{\ensuremath{S_\text{min}}}
\newcommand*{\psnfrac}{\ensuremath{\gamma}}
\newcommand*{\sq}{\ensuremath{\zeta}}
\newcommand*{\halfscatt}{\ensuremath{r}}
\newcommand*{\pvar}{\ensuremath{v}}
\newcommand*{\contrast}{\ensuremath{\mathcal{C}}}
\newcommand*{\loss}{\ensuremath{\ell}}
\newcommand*{\Szbar}{\ensuremath{\bar{\Sz}}}
\newcommand*{\avgSzbarsq}{\ensuremath{\langle\Szbar^2\rangle}}

\usepackage{graphicx}

\usepackage{hyperref}
\hypersetup{
    pdfstartview={FitH},
    pdftitle={},
    pdfauthor={Ian D. Leroux}}
\ifpdf
\fi

\begin{document}

\title{Unitary Cavity Spin Squeezing by Quantum Erasure}

\author{Ian D. Leroux}
\altaffiliation[Present Address: ]{Department of Physics and Astronomy, Aarhus University, DK-8000 Aarhus C, Denmark}

\author{Monika H. Schleier-Smith}
\altaffiliation[Present Address: ]{Max-Planck-Institut f\"{u}r Quantenoptik and Ludwig-Maximilians-Universit\"{a}t, Schellingstra{\ss}e 4, 80799 München, Germany}

\author{Hao Zhang}

\author{Vladan Vuleti\'{c}}
\affiliation{
Department of Physics, MIT-Harvard Center for Ultracold Atoms,
and Research Laboratory of Electronics, Massachusetts Institute of Technology,
Cambridge, Massachusetts 02139, USA}

\date{\today}

\begin{abstract}
Deterministic light-induced spin squeezing in an atomic gas is limited by photon shot noise or, equivalently, by atomic state information escaping with the light field mediating the effective atom-atom interaction.
We show theoretically that the performance of cavity spin squeezing [M.H. Schleier-Smith, I.D. Leroux, and V. Vuleti\'{c}, Phys. Rev. A \textbf{81}, 021804(R) (2010)] can be substantially improved by erasing the light-atom entanglement, and propose several methods for doing so.
Accounting for light scattering into free space, quantum erasure improves the scaling of cavity squeezing from $S^{-1/2}$ to $S^{-2/3}$, where $S$ is the total atomic spin.
\end{abstract}
\maketitle

\section{Introduction}

Squeezed spin states \cite{Kitagawa1993} are among the simplest many-body entangled states to describe and characterize, since their quantum correlations appear as an improved signal-to-noise ratio in certain measurements.
This improvement makes squeezed spin states potentially useful for precision metrology beyond the standard quantum limit (SQL) that is set by the quantum fluctuations of independent particles.
In particular, spin squeezing might increase the stability of atomic clocks, magnetometers, and other measurements based on atom interferometry \cite{Wineland1992,Wineland1994,Louchet-Chauvet2010,Leroux2010:clock}.
Many approaches to spin squeezing have been proposed~\cite{Wineland1992,Kuzmich1998,Sorensen1999,Sorensen2001,Andre2002,Bouchoule2002,Zhang2003,Stockton2004,Hammerer2004,Madsen2004,Takeuchi2005,Genes2006,Meiser2008,Saffman2009,Schleier-Smith2010:dynsq,Schleier-Smith2011:err,Tasgin2011} and a number of them have been demonstrated experimentally, including atomic absorption of squeezed light~\cite{Hald1999}, entangling gates in an ion trap~\cite{Meyer2001}, projection by quantum non-demolition (QND) measurement~\cite{Appel2009,Schleier-Smith2010:msmt,Chen2011}, atom-atom collisions~\cite{Esteve2008,Gross2010,Riedel2010}, and light-mediated interaction between distant atoms in an optical resonator~\cite{Leroux2010:dynsq,Leroux2011:err}.
This last method, cavity squeezing~\cite{Schleier-Smith2010:dynsq,Schleier-Smith2011:err}, has generated the strongest spin squeezing to date, with 5.6 dB observed (no noise subtracted) and 10 dB inferred (detection noise subtracted)~\cite{Leroux2010:dynsq}.

Cavity squeezing relies on the off-resonant interaction between an ensemble of atoms and a light field circulating in an optical resonator cavity~\cite{Schleier-Smith2010:dynsq}.
The atoms' state-dependent index of refraction modifies the cavity resonance frequency.
If the cavity is driven by a probe laser, the atom-induced resonance frequency shift changes the optical power circulating in the cavity, modifies the AC Stark shift it imparts to the atoms, and thus affects the phase of the atomic state.
This phase shift, which depends on the states of all atoms in the ensemble, introduces correlations between them and produces a squeezed state of the total atomic spin.
In the cavity squeezing scheme analyzed in Ref.~\cite{Schleier-Smith2010:dynsq}, the atomic spin is entangled with the outgoing light field so that an outside observer can gain information about the atomic state.
While that information can be used to reduce the variance of a transverse spin component~\cite{Kuzmich1998,Kuzmich2000,Takano2009} and hence to perform conditional spin squeezing by QND measurement~\cite{Appel2009,Schleier-Smith2010:msmt,Chen2011}, it is ignored in unconditional cavity squeezing and thus causes non-unitary evolution in the spin subspace.
Equivalently, the AC Stark shift fluctuations associated with the probe photon shot noise cause an undesirable growth of the spin uncertainty region, reducing the achievable cavity squeezing.

\begin{figure}
  \centering
  \includegraphics[width=\linewidth]{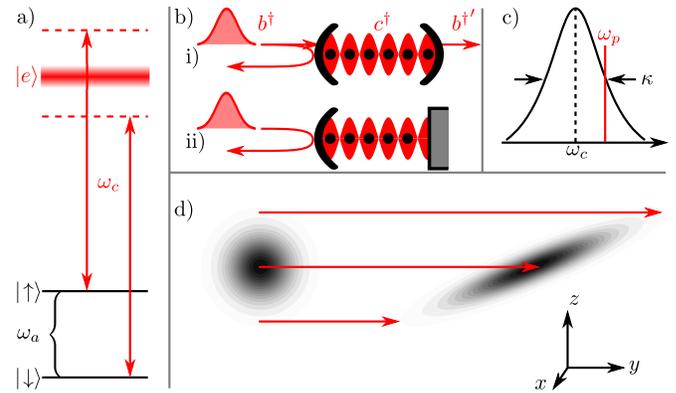}
  \caption{(Color online) Setup for cavity squeezing.
    (a) Atoms with two pseudospin states $\upk,\downk$ optically coupled to an excited state $\exck$.
    (b) The atoms are uniformly coupled to an optical cavity, either symmetric (i) or one-sided (ii), whose resonance frequency $\ocav$ lies between the two optical transition frequencies.
    (c) The cavity is driven by a pulse detuned from its resonance frequency, so that the intracavity intensity depends on the atom-induced cavity frequency shift.
    (d) The resulting AC Stark shift imparts an \Sz-dependent phase shift to the atoms, shearing the uncertainty distribution of a coherent spin state on the Bloch sphere of total atomic spin $\vc{S}$.}
  \label{fig:setup}
\end{figure}

In this paper we propose schemes that suppress the effects of photon shot noise, disentangle the outgoing light field and the atomic spin, and ideally result in unitary evolution in the spin subspace.
We show that the photon shot noise suppression can be achieved either by creating effective photon number states using high-efficiency photodetectors and classical feedback or by a spin echo technique for canceling quantum noise.
Practically, this proposal shows that the performance of unconditional light-induced squeezing can be improved well beyond the limit set by photon shot noise or light-atom entanglement.
Fundamentally, it shows that one can engineer light-mediated interactions between distant atoms without leaving a trace of the atomic state in the light field.

A related idea has recently been proposed~\cite{Trail2010} in a modification of the free-space scheme put forward by Takeuchi \emph{et al.} for spin squeezing by two-fold light-atom interaction \cite{Takeuchi2005}.
The modified scheme of Ref.~\cite{Trail2010} can produce exponential squeezing for short times, but is ultimately limited by light scattering into unobserved free-space modes.
The near-unity single-atom cooperativity readily available in optical resonators~\cite{Tanji-Suzuki2011} reduces the rate of this scattering relative to the squeezing, and allows the scheme we present here to achieve substantially greater spin squeezing for a given atom number than any free-space technique.

In Sec.~\ref{sec:ideal} we analyze cavity squeezing in a one-sided cavity where the coupling of the intracavity light field to the continuum of external light modes is treated exactly in the input-output formalism~\cite{Collett1984,Gardiner1985}.
For an input light field prepared in a near-monochromatic photon number state no information about the atomic state is contained in the outgoing light field and the spin dynamics are therefore unitary.
We show that the evolution of the collective atomic spin is then described by the one-axis twisting Hamiltonian introduced by Kitagawa and Ueda \cite{Kitagawa1993}. 
In Sec.~\ref{sec:lightnoise} we introduce the effects of non-ideal input light states, and in Sec.~\ref{sec:schemes} we use the results to evaluate several practical implementation strategies: one using a spin-echo technique to erase the phase information from the outgoing light field, one using a squeezed input light field, and one using high-efficiency photodetectors to generate an effective photon number state of the light field.
In Sec.~\ref{sec:scatt} we consider the unavoidable effects of light scattering into free space by the atomic ensemble, deriving the associated limit on cavity squeezing with quantum erasure.

\section{Unitary Cavity Squeezing Using an Input Photon Fock State} \label{sec:ideal}

In Ref.~\cite{Schleier-Smith2010:dynsq} the authors analyze cavity squeezing for atoms in a symmetric Fabry-P\'{e}rot cavity driven by a coherent field tuned to the slope of the resonance [Fig.~\ref{fig:setup}(b)(i)].
The performance of the scheme is limited by the entanglement of the atomic spin variables with the light field leaving the resonator.
If the information in the light field is not used, then it must be traced over, leading to decoherence of the atomic state.
This adds spin uncertainty which ultimately limits the attainable amount of spin squeezing.

That this decoherence can be eliminated is most easily seen in a setup with a one-sided cavity [Fig.~\ref{fig:setup}(b)(ii)].
Provided that the cavity is lossless, all incident light is reflected, and nothing about the intracavity dynamics can be inferred from the reflected power.
If the near-monochromatic input light is in a photon number state rather than a coherent state, then any phase shift imparted to the reflected light cannot be measured either.
With no way for an outside observer to learn the state of the atoms inside the cavity, their dynamical evolution must be unitary, as we show explicitly below.

Briefly, we model the system with an effective Hamiltonian that dispersively couples the atoms' pseudospin degrees of freedom to the optical resonator mode.
After diagonalizing this Hamiltonian, we use it to analyze the evolution of an arbitrary initial pseudospin state under the action of an incident light pulse with definite photon number.
For a sufficiently monochromatic pulse, the system evolves into a product state of the field and pseudospin degrees of freedom, where the transformation of the pseudospin state is generated by the one-axis twisting Hamiltonian~\cite{Kitagawa1993}.

The system we consider consists of $N$ identical atoms uniformly coupled to an optical resonator that, in contrast to the system considered in Ref.~\cite{Schleier-Smith2010:dynsq}, is one-sided (Fig.~\ref{fig:setup}). 
The atoms have two stable states separated in energy by $\hbar \oat$, such as hyperfine or magnetic sublevels, which we label as the pseudospin states $\upk_i$ and $\downk_i$ for the $i$-th atom.
We define total pseudospin operators $2\Sz = \sum_{i=1}^N \left( \upk_i\upb_i - \downk_i\downb_i \right)$, the population difference between the stable states, and $\Splus = \Sx+\ii\Sy = \sum_{i=1}^N\upk_i\downb_i$, $\Sminus = \adj\Splus$.
For simplicity, we consider the case where the two optical transitions $\upk\leftrightarrow\exck$ and $\downk\leftrightarrow\exck$ connecting the stable states to the optically active excited state $\exck$ couple to the resonator with the same single-photon Rabi frequency $2g$.
The cavity resonance frequency $\ocav$ is chosen halfway between the two optical transition frequencies, so that their single-photon detunings $\pm \Delta = \pm\oat/2$ are of equal magnitude and opposite sign.
Within the rotating-wave approximation and neglecting, for now, the scattering of photons into free space by the atoms, the Hamiltonian for the intracavity system is
\begin{align}
  \frac{H_\text{cav}}{\hbar} &= \oat \Sz + \sum_{i=1}^{N} \ocav \exck_i \excb_i + \ocav \nc +\notag \\
  &\qquad + \sum_{i=1}^{N} \left[g \left(\upk_i \excb_i + \downk_i \excb_i\right)\cdag + H.c.\right].
\end{align}
In this equation the first two terms describe the energy of the atoms, the third describes the energy of the cavity mode with photon annihilation operator \cfield{}, and the terms proportional to $g$ describe the coupling of the atoms to the light field.
We are interested in the linear, dispersive regime of atom-field interactions where $\Delta$ is large compared to $g$, to the cavity linewidth $\kappa$, and to the excited-state linewidth $\Gamma$.
Assuming that the intracavity photon number \nc{} remains low enough to keep the excited-state population negligible (i.e. $\avg{\nc} g^2 / \Delta^2 \ll 1$), we adiabatically eliminate the excited state \exck{} to replace the full atom-cavity interaction by an AC Stark shift of the two pseudospin states.
In so doing we also neglect stimulated Raman processes which, under the same set of assumptions, are too far off resonance to be significant.
Using the input-output formalism~\cite{Collett1984,Gardiner1985} to describe the coupling with the field outside the cavity, we obtain the following effective Hamiltonian for the dynamics of the pseudospin and of the light field:
\begin{align} \label{eq:EffHam}
\frac{H}{\hbar} &= \oat \Sz + \ocav \nc
  + \int\dd\omega \omega \nb{\omega} \notag \\
  &\qquad + \Omega \nc \Sz
  + \ii \sqrt{\frac{\kappa}{2 \pi}}
    \int\dd\omega [\bdag{\omega} \cfield - \cdag \bfield{\omega}].
\end{align}
The first three terms of the Hamiltonian account for the energy of the atoms, of the cavity field, and of the field outside the cavity with its continuous spectrum of creation operators \bdag{\omega}.
The fourth term represents the dispersive coupling between the cavity field and the atoms, and may be interpreted either as an AC Stark shift of the atomic levels by the light field or as a modification of the cavity resonance frequency by the atomic index of refraction~\cite{Schleier-Smith2010:dynsq}.
The coefficient $\Omega = 2 g^2 / \Delta$ is both the cavity frequency shift per atomic spin flip and the atomic transition frequency shift per intracavity photon.
The final integral is the coupling between the cavity and the external field through the cavity's partially transmissive input mirror, leading to a damping of the energy stored in the cavity field at rate $\kappa$.

The Hamiltonian $H$ may be exactly diagonalized.
First, note that both $\Sz$ and the total photon number $\nc + \int\dd\omega \nb{\omega}$ are conserved.
Thus, all product states $\ket{m} \otimes \ket{0}$ of an atomic eigenstate $\ket{m}$ of \Sz{} with the electromagnetic vacuum $\ket{0}$ are eigenstates of the full Hamiltonian.
The eigenstates with non-zero photon number require additional labels to specify the spectral distribution of the photons.
To find those eigenstates, we follow Fano's procedure for diagonalizing a discrete state (the cavity mode) coupled to a continuum (the external field)~\cite{Fano1961}, which yields an operator that annihilates a photon in an eigenstate of the total field:
\begin{align}
  \afield{\omega} &=
  \frac{1}{\sqrt{(\omega - \ocav - \Omega \Sz)^2 + \kappa^2 / 4}}
  \left[\ii \sqrt{\frac{\kappa}{2 \pi}} \cfield\right. \notag \\
  &\qquad \left.+ \frac{\kappa}{2 \pi} \pv\int\dd\omega^\prime
    \frac{\bfield{\omega^\prime}}{\omega - \omega^\prime}
    + (\omega - \ocav - \Omega \Sz) \bfield{\omega}\right].
\end{align}
\pv{} is a reminder that the Cauchy principal value of the integral must be taken over the pole at $\omega = \omega^\prime$.
This field operator has the usual commutation relation $\comm{\afield{\omega}}{\adag{\omega^\prime}} = \delta(\omega^\prime - \omega)$ and allows us to rewrite the Hamiltonian (\ref{eq:EffHam}) into the much simpler form
\begin{equation}
  \frac{H}{\hbar} = \oat \Sz + \int\dd\omega \omega \na{\omega}
\end{equation}
in which the first term is just the bare energy of the atoms and the second describes, for a given atomic eigenstate $\ket{m}$, a set of decoupled harmonic oscillators.
The excitations created by the $\adag{\omega}$ operators are photons with amplitudes to be either in the cavity ($\cdag$) or in the outside continuum ($\bdag{\omega^\prime}$), with the intracavity component resonantly enhanced near the atom-shifted cavity resonance frequency $\omega \approx \ocav + \Omega \Sz$.
A one-photon eigenstate can now be generated by acting with the total field raising operator $\adag{\omega}$ on any of the vacuum states,
\begin{equation}
  H \adag{\omega} \ket{m} \otimes \ket{0} =
  \hbar(\oat m + \omega) \adag{\omega} \ket{m} \otimes \ket{0},
\end{equation}
and repeated applications of \adag{\omega} can yield arbitrary $n$-photon states.

We are now equipped to calculate the evolution of the atomic spin state under the action of input light pulses.
In particular, we consider an incident pulse containing exactly $n$ photons incident on the cavity, as described by the initial state
\begin{equation}
  \ket{\Psi_{(-t_0)}} = \ket{\psi_a} \otimes \frac{1}{\sqrt{n!}}
  \left(\int\dd\omega \ee^{\ii \omega t_0} B(\omega) \bdag{\omega}\right)^n
  \ket{0}
\end{equation}
where \ket{\psi_a} is an arbitrary state of the atoms and $B(\omega)$ is the pulse amplitude spectrum.
We have written this initial state explicitly as a product state of the atoms and field, emphasizing that the creation operator \bdag{\omega} acts only on the field outside the cavity.
We take the initial time $-t_0$ to be far in the past, before the pulse arrives at the resonator (specifically $t_0 \gg \max \abs{\dd B(\omega) / \dd\omega}^{2/3}$).
Re-expressed in terms of field eigenstates, the initial state is
\begin{align}
  \ket{\Psi_{(-t_0)}} &= \sqrt{\frac{(-\ii)^n}{n!}} \times \\
  &\qquad \times \left(\int\dd\omega \ee^{\ii (\omega t_0 - \Phi_\omega)} B(\omega) \adag{\omega}\right)^n \ket{\psi_a} \otimes \ket{0} \notag
\end{align}
where we have introduced an atom-dependent operator
\begin{equation}
  \Phi_\omega =
  \arctan \left(\frac{\kappa / 2}{\omega - \ocav - \Omega \Sz}\right)
  - \frac{\pi}{4}
\end{equation}
which represents the phase lag of the intracavity field's response to an external drive at frequency $\omega$.
In this form, all components of \ket{\Psi} evolve with known frequency, so it is straightforward to find the final state at $+t_0$, long after the light pulse has reflected from the cavity.
Writing this final state in terms of operators that act separately on the field (\bdag{\omega}) and on the atoms ($\Sz$), we see that it is, in general, entangled:
\begin{align}
  \ket{\Psi_{(+t_0)}} &= \frac{(-\ii)^n \ee^{-2 \ii \oat t_0
      \Sz}}{\sqrt{n!}} \times \\
  &\qquad \times \left(\int\dd\omega \ee^{-\ii (\omega t_0 + 2 \Phi_\omega)}
    B(\omega) \bdag{\omega}\right)^n \ket{\psi_a} \otimes \ket{0}.\notag
\end{align}
However, for a near-monochromatic input pulse centered on a frequency \oprobe{} with a bandwidth much less than $\kappa$, the atomic operator $\Phi_\omega$ is approximately $\Phi_\oprobe$ over the spectrum of the pulse and the state factorizes into
\begin{align}
  \ket{\Psi_{(+t_0)}} &= \ee^{-2 \ii (n \Phi_\oprobe + \oat t_0 \Sz)}
    \ket{\psi_a} \otimes \\
  &\qquad \otimes \frac{(-\ii)^n}{\sqrt{n!}}
  \left(\int\dd\omega B(\omega) \ee^{-\ii \omega t_0} \bdag{\omega}\right)^n \ket{0}.\notag
\end{align}
In this limit of an incident monochromatic $n$-photon Fock state the final field state is independent of the atomic state (the pulse has simply been reflected by the one-sided cavity), while the atomic state has been transformed by the unitary operator
\begin{equation}
  U_n=\ee^{-2 \ii (n \Phi_\oprobe + \oat t_0 \Sz)}.
\end{equation}
The dynamics of interest in the spin subspace are encoded in the nonlinear operator $\Phi_\oprobe$.
When the pulse frequency is tuned to the slope of the Lorentzian cavity resonance, $\oprobe = \ocav + \kappa / 2$, and provided that $N \Omega \ll \kappa$ so that the atoms do not shift the cavity resonance frequency by a large fraction of the cavity linewidth, we can expand $\Phi_\oprobe$ to second order in the spin rotation angle $\pphase = 2 \Omega / \kappa$ imparted by a single incident photon.
We find
\begin{equation}
  U_n = \ee^{-\ii \left(2 \oat t_0 \Sz + n \pphase \Sz + \frac{n}{2} \pphase^2 \Sz^2\right)},
\end{equation}
where the terms linear in \Sz{} generate a precession of the atomic pseudospin and the term quadratic in \Sz{} generates a shearing or \Sz{}-dependent rotation.

To isolate the shearing term it is convenient to apply this transformation twice, separated by a $\pi$ pulse on the atoms which inverts \Sz{}.
The overall effect is the two-pulse transformation
\begin{equation}
  \label{eq:Urhomu}
  U_{\rho,\mu} = \ee^{-\ii \left(\rho \Sz + \frac{1}{2} \mu \Sz^2 \right)}
\end{equation}
obtained by the sequence $U_{n_1}$--$\pi$--$U_{n_2}$ with photon numbers $n_1$ and $n_2$.
Here $\rho = (n_2 - n_1) \pphase$ is the phase rotation angle and $\mu = (n_2 + n_1) \pphase^2$ is the strength of the shearing action expressed as an atomic phase shift per unit change in \Sz{}.
In the ideal case where the incident photon number is identical in the two pulses, $n_1=n_2$, we find the one-axis twisting transformation $U_{0,\mu} = \ee^{-\ii \mu \Sz^2 / 2}$ considered by Kitagawa and Ueda~\cite{Kitagawa1993}.
Their results for the spin squeezing achievable by this transformation are reviewed in Appendix~\ref{app:oat}.

Thus we find that the two-fold interaction of an $n$-photon pulse with the atom-cavity system can realize the $\Sz^2$ one-axis twisting Hamiltonian, at least when scattering into free space is ignored (see Sec.~\ref{sec:scatt}).
The unitary evolution in the spin subspace can be understood as a consequence of the absence of photon shot noise, or equivalently, of the absence of phase information in the reflected light that would reveal the atomic state.
While the same transformation can be accomplished with a single input pulse followed by a photon-number-dependent rotation of the atomic spin, it is easier to implement using the double-pulse sequence, as we shall see below.

\section{Effects of Uncertain Photon Number}
\label{sec:lightnoise}

For input pulses with uncertain photon number, we can find the expectation value of an atomic observable by averaging over the possible photon numbers of the input state.
In general, such an average will depend on all moments of the photon number distribution.
However, an important special case arises when the phase per photon $\pphase$ is small and the photon numbers are large such that the distributions for $n_1$ and $n_2$ approach Gaussians.
In the common case where the fluctuations in the photon number difference $(n_2 - n_1)$ have a variance which
scales with the total photon number, the rotation angle variance can be expressed as $\var\rho = \psnfrac \avg\mu$, with \psnfrac{} a proportionality constant which is 1 for the case of independent photon shot noise fluctuations on $n_1$ and $n_2$.
The variance of the shearing $\var{\mu}$, meanwhile, is higher-order in \pphase{} and vanishes in the limit we are considering, so we will not distinguish between $\mu$ and its expectation value $\avg\mu$ hereafter.
In this regime the squeezing is determined only by the total number of atoms $2S$, the shearing $\mu$ set by the mean total photon number $\avg{n_1+n_2}$, and the ratio \psnfrac{} of rotation uncertainty to shearing set by the variance of the photon number difference $n_1-n_2$ between the two pulses.
For large $S$, an initial atomic pseudospin polarized along $\unitx$, $\mu \ll 1$, and $\avg{\rho}=0$, the final spin expectation values are approximated as
\begin{align}
  \avg\Sx &= S \ee^{-\frac{1}{2} \pvar}, \\
  \avg\Sy &= \avg\Sz = 0, \\
  \avg{\Sy^2} &= \frac{S}{2} \Bigl[1 + S \bigl(1 - \ee^{-2 \pvar}\bigr)\Bigr], \\
  \avg{\Sz^2} &= \frac{S}{2}, \\
  \avg{\Sy \Sz + \Sz \Sy} &= 2 \avg{\Sz^2} \mu \avg{\Sx} = S^2 \mu \ee^{-\frac{1}{2} \pvar} \label{eq:SySz},
\end{align}
where $\pvar = \psnfrac \mu + \mu^2 \avg{\Sz^2}$ is the characteristic phase variance resulting from both the uncertainty on the photon number and from the shearing.
The mean spin length \avg{\Sx} is reduced as the phase broadening \pvar{} wraps the uncertainty region around the Bloch sphere.
The transverse variance \avg{\Sy^2} is initially the projection noise $S/2$, then grows as the phase variance scaled by the length of the Bloch vector $S^2\pvar$, before saturating near $S^2/2$ when the uncertainty distribution has completely wrapped around the Bloch sphere.
The cross correlation $\avg{\Sy \Sz + \Sz \Sy}$ is the product of the variance of \Sz{}, the phase change $\mu$ per unit \Sz{}, and the change in \Sy{} per unit phase.
Using Eq.~(\ref{eq:Smin}) to compute the minimum transverse spin variance $\var{\Smin}$ and comparing this variance to the reduced mean spin length \avg{\Sx} gives the metrological squeezing parameter~\cite{Wineland1992,Wineland1994}
\begin{equation}
  \sq = \frac{2 S \var{\Smin}}{\avg{\Sx}^2}.
\end{equation}
$\sq^{-1}$ is the squared signal-to-noise ratio, relative to the standard quantum limit, of a measurement of the total pseudospin's direction.

\begin{figure}
  \centering
  \includegraphics{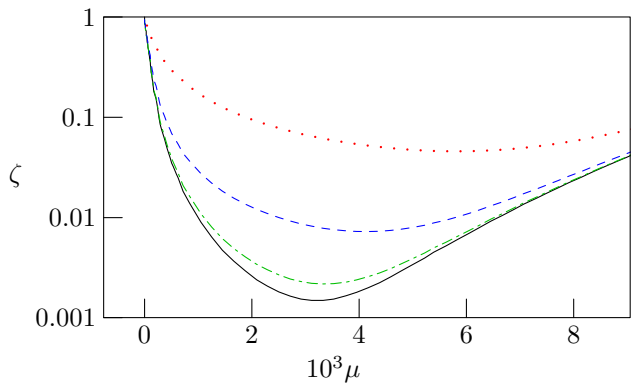}
  \caption{(Color online) Cavity squeezing for $S=10^4$ and varying degrees of photon shot noise suppression: none ($\psnfrac = 1$, dotted red), 90\% ($\psnfrac = 0.1$, dashed blue), 99\% ($\psnfrac = 0.01$, chain dotted green), and complete ($\psnfrac = 0$, solid black).
  In this figure, noise due to photon scattering into free space is ignored, corresponding to a cavity with single-atom cooperativity parameter $\eta \gg 1$ (see Sec.~\ref{sec:scatt}).}
  \label{fig:gaussian-squeezing}
\end{figure}
Figure~\ref{fig:gaussian-squeezing} shows the squeezing parameter calculated as a function of shearing strength $\mu$ for $S=10^4$ and residual fractions of photon shot noise \psnfrac{} of 1, 0.1, 0.01, and 0.
In general the squeezing parameter first decreases as $\sq \approx 2 \psnfrac / (S \mu)$ under the action of the shearing before rising again as $\sq \approx S^2 \mu^4 / 24$ as the curvature of the Bloch sphere deforms the uncertainty region.
In the limit $\psnfrac \rightarrow 0$ we recover the ideal squeezing from perfect one-axis twisting as considered in Ref.~\cite{Kitagawa1993} and Appendix~\ref{app:oat}: $\sq \approx (S \mu)^{-2} + S^2 \mu^4 / 24$, leading to an optimum squeezing that scales as $\sq \approx 12^{2/3} / (8 S^{2/3})$.

Even if the input field contains a definite photon number, imperfections in the resonator can introduce optical loss.
Since the loss of photons is a random process, it reintroduces noise on the number of photons which interact with the atoms, and thus decoherence.
Such optical loss processes can quite generally be understood as coupling the cavity to a second continuum of modes into which the photons are scattered.
For example, we might imagine a second continuum of modes $\bfield\omega^\prime$ behind the right-hand cavity mirror of Fig.~\ref{fig:setup}(b)(i) and then allow this mirror to be partly transparent.
Since the cavity field \cfield{} couples to a weighted sum of the incident field mode \bfield{\omega} and the various independent unobserved modes $\bfield\omega^\prime$ we can, following Fano, identify a single linear combination of all the field modes at a given frequency that couples maximally to the cavity mode and one or more orthogonal combinations (as many as there are unobserved fields) that do not couple to the cavity mode at all~\cite{Fano1961} (see Appendix~\ref{sec:sagnac} for a practical consequence of this separation into coupled and uncoupled modes).
If the average fraction lost of the input pulse at one half-linewidth detuning is $\loss \ll 1$, then the probability of an incident photon being in one of these uncoupled modes and failing to interact with the cavity mode is $\frac{1}{2} (1 - \sqrt{1   - 2 \loss}) \approx \loss / 2$, yielding a binomial distribution for the cavity-coupled photons whose variance is $\psnfrac \approx \loss / 2$ times photon shot noise.
In other words, the noise in the photon number difference $n_2 - n_1$ to which the squeezing is sensitive is just the shot noise of the lost photons.

\section{Practical Photon Shot Noise Suppression Schemes} \label{sec:schemes}

In this section we show how to achieve reduced photon shot noise $\psnfrac<1$ using several schemes of practical interest.

\subsection{Spin Echo Quantum Eraser for Coherent Input Pulses}

\begin{figure}
  \centering
  \includegraphics{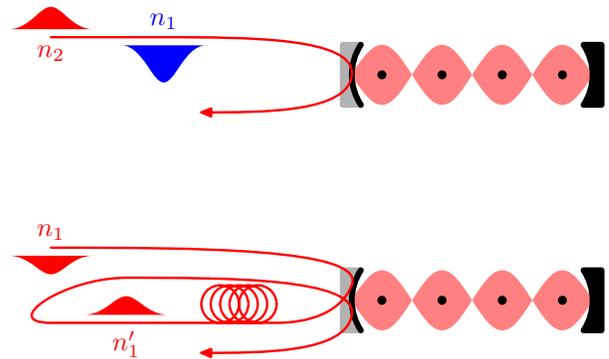}
  \caption{(Color online) Cavity squeezing with coherent pulses.
    Previous experimental demonstrations of cavity squeezing~\cite{Leroux2010:dynsq} used two separate coherent pulses with statistically independent photon numbers $n_1$ and $n_2$ (top).
    If, instead, a single coherent pulse is used for both interactions with the cavity (bottom),
    the two photon numbers $n_1$ and $n_1^\prime$ differ only by the optical losses between the two interactions
    and the effect of their correlated fluctuations can be canceled out by a spin-echo sequence.}
  \label{fig:double-pulses}
\end{figure}

If the input consists of two independent coherent pulses (Fig.~\ref{fig:double-pulses}, top), $\var{(n_2 -  n_1)} = \avg{n_2} + \avg{n_1}$ and $\psnfrac = 1$.
In this case we recover the scaling obtained by Takeuchi \emph{et   al.}~\cite{Takeuchi2005} for their polarization-feedback scheme in free space: $\sq \approx 2 / (S \mu) + S^2 \mu^4 / 24$, with the best squeezing, obtained for a shearing parameter $\mu \approx 12^{1/5} / S^{3/5}$, scaling as $\sq \approx 5 / (384^{1/5} S^{2/5})$.
Note that these results do not take into account free-space scattering, which can be ignored only if a cavity with cooperativity $\eta>1$ is used (see Sec.~\ref{sec:scatt}).

However, if the same coherent pulse is reused for both interactions by storing it in an optical delay line while the $\pi$ pulse is applied to the atoms (Fig.~\ref{fig:double-pulses}, bottom), then ideally $n_2 = n_1$ and $\rho=\psnfrac = 0$.
Information about \Sz{} is encoded in the optical phase shift of the coherent pulse after its first reflection from the cavity, but this is erased during the second reflection, which applies the opposite phase shift since \Sz{} has changed sign in the interim.
Thus the light is disentangled from the atoms at the end of the sequence and the overall evolution of the atoms is unitary.
In a real implementation there will be losses in the delay line used to store the pulse between its two interactions with the cavity, such that $n_2$ will not be precisely equal to $n_1$ and $\rho$ will not exactly vanish.
If a fraction $\loss$ of the photons is lost between the two pulses, then $\var{(n_2 - n_1)} = \avg{n_1} \loss \approx (\avg{n_2} + \avg{n_1}) \loss / 2$, where the second approximation holds for small losses, giving $\psnfrac \approx \loss/2$.
Again, the squeezing is limited by the shot noise of the lost photons.

\subsection{Squeezed Input Pulses}

One demonstrated approach to improving optical atom detection for a given number of photons is to use a squeezed state of the input light field~\cite{Wolfgramm2010}.
We therefore consider the effect of incident pulses whose fluctuations in-phase with the coherent amplitude $\alpha$ are squeezed so as to reduce intensity noise in the cavity.
The photon number variance of such pulses can be parametrized as~\cite{Yuen1976,Loudon1987} $\var{n} = \abs\alpha^2 \ee^{-2 s} + 2 \sinh^2 s \cosh^2 s$ where $\alpha$ is the coherent amplitude of the pulse.
The usual optical squeezing parameter $s$ is related to the shot noise suppression factor by $\psnfrac = \ee^{-2 s} + (2 \sinh^2 s \cosh^2 s) / \avg{n_2 + n_1}$.
Note that $\psnfrac$ does not decrease monotonically with $s$ but reaches a minimum which improves with photon number as $\psnfrac \sim (n_2 + n_1)^{-1/3}$, because squeezing reduces the fluctuations in a quadrature of the incident electric field rather than in its magnitude.
Delivering such squeezed light to the cavity in the presence of optical losses remains experimentally challenging, but sources providing \unit[9]{dB}-squeezed light to gravitational observatories have been demonstrated~\cite{Vahlbruch2010}, so that suppression factors of order $\psnfrac\sim10^{-1}$ may be attainable by this approach.

\subsection{Generation of Effective Fock States by Measurement} \label{sec:Fock}

\begin{figure}
  \centering
  \includegraphics{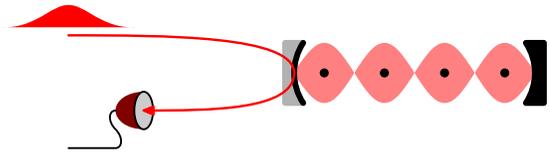}
  \caption{(Color online) By counting photons after their interaction with the cavity,
    the squeezing light pulse can be projected onto a definite photon number state.
    With fast classical feedback it can even be steered to a predetermined number state
    for unconditional photon shot noise suppression.}
  \label{fig:fock-projection}
\end{figure}

Another approach to suppressing photon shot noise relies on the conservation of total photon number in this scheme: every photon sent onto the cavity must leave it and, for a single-ended cavity with negligible loss, every photon leaves in the same reflected mode.
This identity between photon numbers of the input and output fields allows an effective input Fock state to be produced using high-efficiency photon counters and classical feedback.
Such feedback-generated Fock states were studied in the early days of light squeezing research~\cite{Machida1986} but have not seen wide use because, in the absence of a QND photodetector, the photon Fock state is destroyed in the very detection process that generates it.
However, since the spin-squeezing setup does not change the photon number we may place it inside the feedback loop, before the photodetector (Fig.~\ref{fig:fock-projection}), thus sidestepping this difficulty.

In the simplest scheme, a coherent laser pulse is sent onto the cavity and the reflected light is collected by a photon counter. 
If a perfect photon counter detects $n$ photons in the light pulse reflected from the cavity, then the system is projected into the state obtained for an incident $n$-photon Fock state.
The photon-counting measurement has removed the uncertainty on the energy of the incident light pulse and has destroyed the complementary information in the phase of the light field, which would have revealed the atomic state via the cavity frequency shift.  Conditioned on the reflected photon number, the atomic dynamics are unitary: although the evolution of the spin state depends \emph{a priori} on the uncertain photon number in the incident pulse, \emph{a posteriori} the experimenter knows exactly which unitary operation was performed by the light pulse whose photon number was measured.
Note that no useful information can be obtained from the photon arrival times at the counter.
Since the spectrum of the incident light must be much narrower than the cavity linewidth, the arrival time of the photons has a Fourier-limited uncertainty much larger than the cavity lifetime and one cannot determine whether a photon entered the resonator or merely bounced off the input mirror.

Rather than contenting oneself with conditional unitary evolution, one can deterministically generate the equivalent of an input Fock state by applying direct feedback to the incident light. 
One must merely count reflected photons and switch the light source off once some target photon number $n$ has been reached.

Finite photodetector quantum efficiency will introduce an uncertainty on the number of input photons for a given detected photon count.
The residual photon shot noise obtained by this technique will therefore be at best $\psnfrac = 1-Q$, where $Q$ is the quantum efficiency of the photodetector.
Note that a transmission-based QND measurement of the atomic spin as used in Ref.~\cite{Schleier-Smith2010:msmt} can squeeze initially as $\sq \approx 2 / (S \mu Q)$, which is slower than the $\sq \approx 2 \gamma / (S \mu)$ of cavity squeezing for any finite quantum efficiency.
Even a perfect photon-shot-noise-limited phase measurement of the light reflected from a one-sided cavity could squeeze only as $\sq \approx 1 / (4 S \mu Q)$, so that for $Q\gtrsim\unit[85]{\%}$ conditional squeezing by measurement still proceeds more slowly than cavity-feedback squeezing using the same photodetector to suppress photon shot noise fluctuations.

\section{Effects of Scattering Into Free Space} \label{sec:scatt}

So far, we have neglected scattering of photons into free space.
Like atom loss in collisional squeezing of Bose-Einstein condensates~\cite{Sinatra2011}, such photon loss degrades the performance of light-induced spin-squeezing schemes~\cite{Hammerer2004,Madsen2004} unless a suitable level scheme is used to avoid its effects~\cite{Saffman2009}.
In the simple and symmetric model we consider, the average number of photons scattered into free space per atom in the ensemble is given by
\[
  2\halfscatt = (n_1 + n_2) \frac{\pphase}{2} \frac{\Gamma}{\Delta} = \frac{\mu}{2 \eta}
\]
where the numbers of photons Raman- and Rayleigh-scattered are equal to each other and to $\halfscatt$.
The scattering depends only on the shearing $\mu$ and the cavity cooperativity $\eta = 4 g^2 / (\kappa \Gamma)$, so that for any finite single-atom cooperativity $\eta$ scattering into free space is inescapable at any detuning $\Delta$.
Scattering leads directly to atomic decoherence by revealing the state of certain atoms to a hypothetical observer outside the cavity and, in the case of Raman scattering, by randomly flipping some spins in the ensemble.
Here we apply the treatment of these effects given in Ref.~\cite{Schleier-Smith2010:dynsq} to the case of unitary cavity feedback.

All photons scattered into free space reveal the internal state of the scattering atom to a hypothetical observer.
For Raman scattering, the state is encoded in the frequency of the scattered light.
For Rayleigh scattering, it is encoded in the phase of the scattered field, because the laser detuning from resonance has opposite sign for the two spin states~\cite{Uys2010}.
Any atom which scatters a photon into free space therefore acquires an unknown phase, entangled with the information lost in the scattered field and uncorrelated with that of the other atoms in the ensemble.
The mean length of the Bloch vector is thus reduced from $S$ by a factor of $\contrast = \ee^{-2 \halfscatt}$ corresponding to the fraction of atoms which have scattered no photons.

Raman scattering, in addition, modifies the relative population of $\upk$ and $\downk$, and forces us to distinguish between the spin component $\Sz$ found at the end of the squeezing and its time-averaged value $\Szbar$ during the squeezing process.
The distribution of $\Sz$ is unaltered by the Raman scattering, since it already corresponded to the sum of independent random $\pm1/2$ contributions from the uncorrelated atoms in the initial state.
But since spins which flip partway through the squeezing pulse contribute less to the time average, $\avgSzbarsq$ is reduced to $(1 - 2 \halfscatt / 3) S / 2$, to leading order in $\halfscatt$.
Since it is $\Szbar$ which sets the average atom-induced shift of the light intensity inside the cavity, and thus the phase shift imparted to the atoms by the Stark effect, the phase variance in turn is reduced to $\pvar^\prime \approx \psnfrac \mu + \mu^2 \avgSzbarsq$.
Similarly, the factor of $\avg{\Sz^2}$ in the $\Sy$--$\Sz$ correlation [Eq.~(\ref{eq:SySz})]---which expressed the correlation between the $\Sz$ found at the end of the squeezing and the atom-induced change to the light shift that modifies $\Sy$---becomes $\avg{\Sz \Szbar} \approx (1 - \halfscatt) S / 2$, again to leading order in $\halfscatt$.

\begin{figure}
  \centering
  \includegraphics{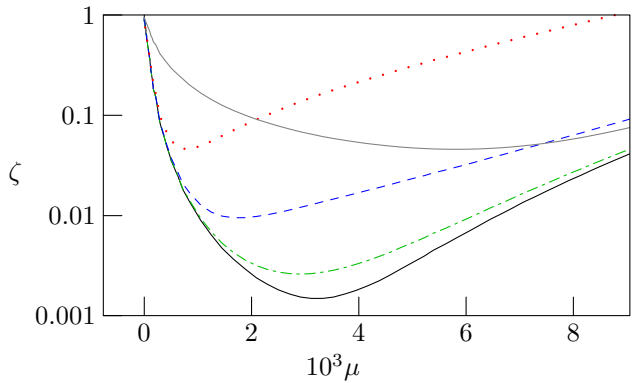}
  \caption{(Color online) Cavity squeezing for $S=10^4$ with input pulses of definite photon number for a perfect cavity (solid black), and for finite single-atom cooperativities $\eta=1$ (chaindotted green), $\eta=0.1$ (dashed blue) and $\eta=0.01$ (dotted red).
    For reference, the gray curve shows squeezing for a perfect cavity ($\eta\rightarrow\infty$) without photon shot noise suppression ($\psnfrac=1$).}
  \label{fig:squeeze-w-scatt}
\end{figure}

Combining these effects, we find an adjusted set of spin moments
\begin{align}
  \avg\Sx &= S \contrast \ee^{-\frac{1}{2} \pvar^\prime},  \\
  \avg\Sy &= \avg\Sz = 0, \\
  \avg{\Sy^2} &= \frac{S}{2},
  \Bigl[1 + S \contrast^2 \bigl(1 - \ee^{-2 \pvar^\prime}\bigr)\Bigr], \\
  \avg{\Sz^2} &= \frac{S}{2}, \\
  \avg{\Sy \Sz + \Sz \Sy} &= S^2 (1 - \halfscatt) \contrast \mu
  \ee^{-\frac{1}{2} \pvar^\prime}.
\end{align}
Note that photon number fluctuations due to atomic absorption can be neglected because, while the fraction of atoms which scatter a photon is fixed by $\mu$ and $\eta$, the fraction of photons scattered vanishes in the large-photon-number limit considered here.
The resulting squeezing is plotted in Fig.~\ref{fig:squeeze-w-scatt} for different cavity cooperativities $\eta$ and for perfect photon shot noise suppression ($\psnfrac=0$).

For the weak- and moderate-coupling regimes where scattering is the dominant limitation on squeezing ($\eta < 1$), we find $\sq \approx (S \mu)^{-2} + \mu / (3 \eta)$.
As in the ideal case, the squeezed variance is initially suppressed by the square of the squeezing parameter, but the noise from scattering into free space adds a variance which scales linearly with the shearing.
This leads to an optimum squeezing $\sq \approx 6^{1/3} / 2 (S \eta)^{2/3}$ for a shearing parameter $\mu \approx (6 \eta)^{1/3} / S^{2/3}$.
Note that $4 S \eta$ corresponds to the resonant optical depth of the atomic ensemble probed through the cavity, and that the achievable squeezing therefore scales as optical depth to the $-2/3$.
This is the same scaling reported by Trail \emph{et al.} for their analogous polarization-based spin-squeezing scheme~\cite{Trail2010}.
The dashed curve of Fig.~\ref{fig:squeeze-w-scatt} shows the squeezing achievable with photon shot noise suppression in a setup otherwise similar to that used in Ref.~\cite{Leroux2010:dynsq}, using $2S=2\times 10^4$ atoms of $\Rb$ ($\Gamma / \abs\Delta = 1.8\times 10^{-3}$) in a resonator with a single-atom cooperativity for the relevant transitions of $\eta=0.1$ so that $\pphase = \eta\Gamma / \abs\Delta = 1.8\times 10^{-4}$.
For a shearing parameter of $\mu=1.8\times 10^{-3}$ corresponding to a photon number of $2.7\times 10^4$ in each incident pulse, the squeezing reaches \unit[20]{dB}, a substantial improvement over the \unit[13]{dB} achievable in the same system without photon shot noise suppression.

For the strong-coupling regime $\eta \gg 1$ the curvature becomes significant before scattering can decohere the ensemble, and the ideal squeezing behavior of Sec.~\ref{sec:ideal} is restored.
Note that the scaling of the achievable squeezing with atom number (as $S^{-2/3}$) is the same for finite $\eta$ as it is in this ideal limit of $\eta \rightarrow \infty$.
Once the effect of photon shot noise has been suppressed, the scattering into free space costs only a constant factor in squeezing performance.

\section{Conclusion}

In this paper we have shown how to improve cavity squeezing performance by disentangling the atomic variables from the light field which mediates the interatomic interaction.
We have suggested several ways of doing this, including a spin-echo sequence that erases the phase information in a coherent light pulse, and the use of photodetectors and classical feedback to generate effective Fock states of the input field.
Once the entanglement between atoms and outgoing light field is eliminated, even a moderate cavity cooperativity $\eta \sim 1$ suffices to obtain squeezing performance close to the limit set by the curvature of the Bloch sphere.
This limit, in turn, could be overcome by two-axis counter-twisting~\cite{Kitagawa1993}, which can be realized by alternating periods of one-axis twisting with rotations of the atomic spin~\cite{Liu2011}.
Furthermore, since unitary $\Sz^2$ evolution in combination with rotations suffices, in principle, to implement any unitary map on the Bloch sphere~\cite{Chaudhury2007}, the techniques we have presented could enable the production of non-Gaussian entangled states of ensembles comprising tens of thousands of atoms.
Further studies are needed to determine which states are attainable given realistic experimental imperfections.

I.D.L. acknowledges support from NSERC; M.H.S.-S. acknowledges support from
the Hertz Foundation and the NSF. This work was supported by the NSF, DARPA, and ARO.

\appendix

\section{Squeezing by One-Axis Twisting} \label{app:oat}

This appendix summarizes the results for squeezing by one-axis twisting first obtained by Kitagawa and Ueda~\cite{Kitagawa1993}.
In order to prepare a squeezed state, we begin with the atoms in a totally symmetric but unentangled coherent spin state (CSS) along the \unitx{} axis of pseudospin
\begin{align}
  \ket{\psi_a} &= \left(\frac{\upk + \downk}{\sqrt{2}}\right)^{\otimes N} \\
  &= \sum_{m = -S}^S\sqrt{\frac{1}{2^{2 S}}\binom{2 S}{S + m}}\ket{m}.
\end{align}
The second form explicitly shows the CSS's binomial distribution of \Sz{} eigenvalues.
In this state $\avg{\Sx} = S = N / 2$, $\avg{\Sy} = \avg{\Sz} = 0$, $\avg{\Sy \Sz + \Sz \Sy} = 0$ and $\avg{\Sy^2} = \avg{\Sz^2} = S / 2$.
Afer the transformation $U_{\rho,\mu}$ defined in Eq.~(\ref{eq:Urhomu}), the \Sz{} distribution is unmodified but the phases between \ket{m} levels have acquired a quadratic dependence on $m$.
The expectation values become
\begin{align}
  \avg\Sx &= S \cos (\rho) \cos^{2 S - 1} \left(\frac{\mu}{2}\right), \\
  \avg\Sy &= S \sin (\rho) \cos^{2 S - 1} \left(\frac{\mu}{2}\right), \\
  \avg{\Sy^2} &= \frac{S}{2} \Bigl[1 + \Bigl(S - \frac{1}{2}\Bigr),
  \bigl(1 - \cos (2 \rho) \cos^{2 S - 2} (\mu)\bigr)\Bigr],
\end{align}
and
\begin{align}
  \avg{\Sy \Sz + \Sz \Sy} &= S (2 S - 1) \cos (\rho) \times \notag \\
  &\qquad \times \sin \left(\frac{\mu}{2}\right)
    \cos^{2 S - 2} \left(\frac{\mu}{2}\right).
\end{align}
For $\rho = 0$ the mean spin remains aligned along \unitx{} and the minimum variance transverse to this direction is given by
\begin{equation}
  \var{\Smin} = \frac{1}{2} \left(u_+ -
    \sqrt{u_-^2 + \avg{\Sy \Sz + \Sz \Sy}^2}\right)
  \label{eq:Smin}
\end{equation}
with $u_\pm = \avg{\Sy^2} \pm \avg{\Sz^2}$.
For large $S$ and in the region of significant squeezing $S^{-1} \ll \mu \ll S^{-1/2}$ the squeezing parameter is approximately $\sq \approx (S \mu)^{-2} + S^2 \mu^4 / 24$, decreasing under the action of the shearing until the curvature of the Bloch sphere deforms the uncertainty region.
The best squeezing obtained is $\sq \approx 12^{2/3} / 8 S^{2/3}$ for a shearing parameter $\mu \approx 12^{1/6} / S^{2/3}$.

\section{Simulating a One-Sided Cavity} \label{sec:sagnac}

Throughout this paper we have considered a single-ended cavity.
Many cavity-QED experiments find it convenient to use two-sided Fabry-Pérot resonators.
Such two-sided cavities would mix a Fock state input from one end with vacuum fluctuations admitted through the other end of the cavity, restoring much of the photon shot noise we wish to suppress.
Equivalently, information on the cavity detuning (and hence the atomic state) is available in the ratio of transmitted to reflected photon numbers, and this information leak entails atomic decoherence.
Fortunately, it is possible to convert a symmetric cavity into an effective one-sided cavity using only external optics.

\begin{figure}
  \centering
  \includegraphics[width=\linewidth]{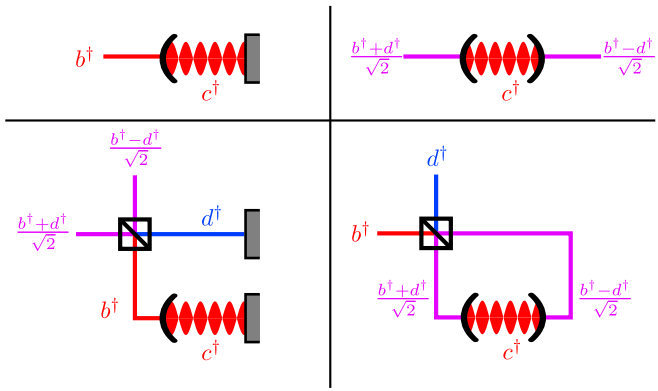}
  \caption{(Color online) Equivalence of one-sided (top left) and two-sided (top right) cavities: the left and right ports of a symmetric Fabry-Pérot resonator can be combined on a beam splitter to isolate the linear combination of the two fields that couples to the intracavity field (bottom right).
    Conversely, the single input of a one-sided cavity can be mixed with an auxiliary mode to yield an effective two-sided resonator (bottom left).}
  \label{fig:cavity-conversion}
\end{figure}

Figure~\ref{fig:cavity-conversion} illustrates the principle of this conversion.
For any given frequency, transverse mode, and polarization, the symmetric cavity couples to two spatially separated input fields (right and left).
Since there is only one cavity mode near the given frequency with the given transverse mode and polarization, it must couple only to some linear combination, labeled here as $b^\dagger$, of the two input fields.
The orthogonal combination $d^\dagger$\ does not couple to the cavity mode at all.
Classically, $b^\dagger$ ($d^\dagger$) corresponds to simultaneous illumination from left and right with phases chosen so as to give constructive (destructive) interference within the resonator.
Enclosing the Fabry-P\'{e}rot in a Sagnac interferometer overlaps the left and right fields at the input, so that with appropriately chosen path lengths the maximally-coupled superposition $b^\dagger$\ is isolated from the uncoupled mode $d^\dagger$.
To an observer looking into the $b^\dagger$\ port of the input beam splitter, the apparatus appears to be a single-ended cavity which couples to no other field modes.
The $d^\dagger$\ input, uncoupled from the cavity, remains available for interferometer stabilization.

\bibliography{shotnoise-cancellation}

\end{document}